\documentstyle[12pt]{article}

\begin{document}

\title{Deformed exterior algebra, Quons and their  Coherent States} 
 \author{} \date{} \maketitle   
\center{\bf M. EL BAZ}\footnote{E-mail address: moreagl@yahoo.co.uk}  
\center{\it Facult\'e des sciences, D\'epartement de Physique, LPT-ICAC, 
\linebreak Av. Ibn Battouta, B.P. 1014, Agdal, Rabat, Morocco} 
\vspace{0.7cm} 
\center{\bf Y. HASSOUNI}\footnote{E-mail address: Y-hassou@fsr.ac.ma}
 \center{\it Facult\'e des sciences, D\'epartement de Physique, LPT-ICAC,
\linebreak Av. Ibn Battouta, B.P. 1014, Agdal, Rabat, Morocco

\vspace{3cm}
\abstract{We review the notion of the deformation of the exterior wedge product. This allows us to construct the deformation of the algebra of exterior forms over a vector space and also over an arbitrary manifold. We relate this approach to the generalized statistics. we study quons, as a particular case of these generalized statistics. We also give their statistical properties.

A large part of the work is devoted to the problem of constructing coherent states for the deformed oscillators. we give a review of all the approaches existing in the literature concerning this point and enforce it with many examples.}

\vspace{1cm}

{\bf Keywords:} Deformation, Wedge product, Algebra of exterior differential forms, Generalized statistics, Quons, Parafermions, Coherent states, Resolution of unity.

\section{Introduction}

The notions of deformation and non-commutativity appeared during the last decade, found their applications in many areas of physics. The quantum groups which are deformations of the usual concepts of groups and (or) algebras (more precisely enveloping algebras) were introduced in the context of the scattering problem and the theory of integrable systems\cite{fadeev, jimbo, drinfeld}. Behind this objects (quantum groups) lies an important structure; the Hopf algebra structure, and the famous Yang Baxter Equation (YBE). 

Q-deformed oscillators were introduced in order to construct q-deformed quantum field theories and quantum mechanics on non-commutative spaces \cite{wess, zumino, schwenk}. We notice that these objects are also used the construction of intermidiate statistics allowing the physical description of some special particles {\it anyons} \cite{leinaas,lerda}. Another mathematical reasoning leading to the same result has been performed in \cite{ando1, ando2, ando3, ando4}. We proved that the deformation of the exterior algebra on a given vector space is related to the YBE\footnote{This is in a sense, the opposite direction usually adopted while constructing the differential calculi on quantum groups see \cite{watts} and references there in}. This is an essential ingredient in the mathematical scheme for constructing generalized statistics. It presents the advantage to be related to non-commutative geometry and the corresponding differential calculus.

Physically these q-oscillators arise in the context of strongly correlated fermions \cite{bonastos1, bonastos2}. We think that by using coherent states and the related tools, one can obtain interesting physical and mathematical results for these q-oscillators \cite{hubbard}. A deep comprehension, then direct physical application could easily arise!

Coherent States (CS) were first introduced by Schrodinger \cite{schro} (though the adjective coherent was not yet used) as quantum states with the closest behaviour to the classical one. Then in the early sixties Glauber \cite{glauber} used these states to describe a "coherent" light beam emitted by lasers in the context of quantum optics. It is he, who used the nomenclature CS for the first time. 

These states present many interesting aspects and properties:

\begin{itemize}

\item These states are related to the harmonic oscillator, and the mathematical structure behind: the Weyl-Heisenberg group. So the CS are generated from the vacuum state by the action of a group-related displacement operator.

\item The states are eigenvectors of the annihilation operator.

\item They are minimum uncertainty states: these states do saturate the Heisenberg uncertainty relation.

\item The states form an overcomplete set of vectors in the associated Hilbert space. This can be seen from the resolution of unity.

\item The states are also stable in time, a coherent state remains coherent.
\end{itemize}

After Glauber's work, many generalizations of the notion of CS appeared. These generalizations deals with one or more of the properties mentioned above. Perelomov and his collaborators for example, have generalized this notion from a group background. So the constructed CS are related to arbitrary lie groups \cite{per72, per} see also \cite{onofri, barut, gilmore}.

Nieto {\it et al} defined CS as states  saturating the Heisenberg uncertainty relations\cite{nieto1, nieto2}.

Coherent states were also constructed for a fermionic harmonic oscillator instead of the bosonic one, for which the original construction was made \cite{ohnuki}.

In \cite{klauder85} the authors defined the minimum set of properties to be imposed on set of states for to be coherent. These properties are the only properties preserved in all the generalizations mentioned above. These are:

\begin{itemize}
\item The states are normalizable.

\item Continue in their label.

\item and provide a decomposition of the unity operator: resolution of unity.
\end{itemize}

For a recent review of all these notions, and a complete list of references see \cite{gazeau}.

On the other hand, an obvious question arise: CS were defined for both bosonic and fermionic harmonic oscillators, how can we define CS related to deformed oscillators? \cite{arik, per96, mine1, mine2}. Despite all the published papers in this direction, many questions remain unanswered concerning this construction. In this paper we are going to address some of these questions. It is divided as follows:

In section 2 we show how to deform the exterior algebra of forms over a vector space $V$. Then in section 3 we extend this to an arbitrary manifold (deformation of the exterior algebra over a manifold). We show how to relate this to the well known (YBE). We also show that such a deformation induces in a natural way, the non-commutativity of the coordinates.

In section 4, we deal with the problem of generalized statistics and how to derive this from the deformed exterior algebra and a previously defined wedge product. In section 5 we deal with a special case of generalized statistics. We obtain this case using special solutions of the (YBE) and also derive some statistical properties of these particles.

Section 6 is devoted to the construction of CS related to the deformed oscillators. We give a review of the different approaches to do so. We give some concrete examples to clarify these ideas.

\section{Deformation of the exterior algebra on a vector space $V$}

In this section, we construct the deformed algebra of exterior forms on a given finite dimensional vector space $V$. This deformation will lead to the Yang-Baxter equation (YBE) when one requires the associativity of the algebra. So, we start by giving a brief review on the definition of the exterior algebra on a given vector space. After that, we proceed to the deformation of this algebra in a special way.

Let $V$ be a vector space over the field $K$ ($K=C$ or $R$) with the finite basis ($e_1,...,e_{dimV})$. The multiplet ($\sigma ^1,...,\sigma ^{dimV}$) is the basis of the dual space $V^*$ of $V$:

\begin{equation}
\sigma ^i (e_j) = \delta ^i _j
\end{equation}

We will show in section 3, where the vector space $V$ is replaced by a finite manifold, that the $\sigma$'s are nothing but one-differential forms over this manifold.

The exterior product of two $\sigma$'s is defined by:

\begin{equation}
\sigma ^i \wedge \sigma ^j = \sigma ^i \otimes \sigma ^j - \sigma ^j \otimes \sigma ^i
\end{equation}

we introduce its deformation, in a natural way as: 
\begin{equation}
\sigma ^i \tilde \wedge \sigma ^j = \sigma ^i \otimes \sigma ^j - \Lambda ^{ij}_{kl} \; \sigma ^k \otimes \sigma ^l
\end{equation}
where $\Lambda ^{ij}_{kl}$ are entries of an arbitrary matrix $\Lambda \epsilon End_K(V^* \otimes V^*)$

Equation (3) can be rewritten in an elegant form:

\begin{equation}
\sigma ^i \tilde \wedge \sigma ^j = \Lambda (\sigma ^i \otimes \sigma ^j) = (E^{ij}_{kl} - \Lambda ^{ij}_{kl}) \sigma ^k \otimes \sigma ^l
\end{equation}
or also
\[
\sigma _1 \tilde \wedge \sigma _2 = ( E_{12} - \Lambda _{12} ) \sigma _1 \otimes \sigma _2
\]
with $ E ^{ij} _{kl} = \delta ^i _k \delta ^j _l$

We notice at this point that the generalization of the exterior wedge product is totally embedded in the matrices $\Lambda$. One can also consider the case where the $\Lambda$'s depend on one or many complex parameters. In any case, the condition to recover the exterior wedge product (2)(at a certain limit) must be saved. In our case this limit is $ \Lambda ^{ij}_{kl} = P^{ij}_{kl} = \delta ^i_l \delta ^j _k $($P$ is the permutation matrix).

Back to our construction, we are in a position to introduce the space of two deformed forms ($\Lambda$-two forms) on $V$. In this context, we notice that the multiplication $\tilde \wedge $ is not anti-symmetric; $\sigma ^i \tilde \wedge \sigma ^j \neq -\sigma ^j \tilde \wedge \sigma ^i$. So the elements $\sigma ^i \tilde \wedge \sigma ^j$ and $\sigma ^j \tilde \wedge \sigma ^i$ are independent {\it a priori}.

We denote by $\Omega ^{(2)}_{\Lambda}(V)$ the space generated by ${\sigma ^i \tilde \wedge \sigma ^j , \;\;i,j=1,...,dim V}$; this is the space of two-deformed forms.

A general $\Lambda$-two form is written:

\begin{equation}
\omega _{\Lambda}^{(2)} = \omega _{ij}\; \sigma ^i \tilde \wedge \sigma ^j
\end{equation}

$\omega _{ij}$ are just complex numbers, and the elements $ \sigma ^i \tilde \wedge \sigma ^j $ belong to $T(V)$ (the tensor algebra of $V$).

In order to construct the other higher order forms, we must define the overlapping between $\otimes$ and $\tilde \wedge$, we put thus:

\begin{eqnarray} 
(\sigma ^i \otimes  \sigma ^j) \tilde \wedge \sigma ^k  &=& \sigma ^i \otimes \sigma ^j \otimes \sigma ^k - \Lambda ^{jk}_{lm} \; \sigma ^i \otimes \sigma ^l \otimes \sigma ^m + \Lambda ^{jk}_{lm} \Lambda ^{il}_{np} \; \sigma ^n \otimes \sigma ^p \otimes \sigma ^m \nonumber \\
\sigma ^i \tilde \wedge ( \sigma ^j \otimes \sigma ^k ) &=& \sigma ^i \otimes \sigma ^j \otimes \sigma ^k - \Lambda ^{ij}_{lm} \; \sigma ^l \otimes \sigma ^m \otimes \sigma ^k + \Lambda ^{ij}_{lm} \Lambda ^{mk}_{np} \; \sigma ^l \otimes \sigma ^n \otimes \sigma ^p
\end{eqnarray}

These relations define the braiding between a $\sigma$ and a $\sigma \otimes \sigma$. This operation is nothing but the transposition one, if one recovers $\wedge$ starting from $\tilde \wedge$ by taking ($\Lambda = P$).

Now, we rewrite eqs (6) in a condensed form as:
\begin{eqnarray}
(\sigma ^1 \otimes \sigma ^2) \tilde \wedge \; \sigma ^3 &=& ( E - \Lambda _{23} +  \Lambda _{23} \Lambda _{12})\; \sigma ^1 \otimes \sigma ^2 \otimes \sigma ^3 \nonumber \\
\sigma ^1 \tilde \wedge \;( \sigma ^2 \otimes \sigma ^3) &=& ( E - \Lambda _{12} +  \Lambda _{12} \Lambda _{23})\; \sigma ^1 \otimes \sigma ^2 \otimes \sigma ^3
\end{eqnarray}
Using these relations we can now introduce the $-three forms$; Basing on equalities (3) and (6), one gets:
\begin{eqnarray}
(\sigma ^1 \tilde \wedge \; \sigma ^2 ) \tilde \wedge \; \sigma ^3 &=& ( E - \Lambda _{23} + \Lambda _{23} \Lambda _{12} - \Lambda _{12} +\Lambda _{12}\Lambda _{23} - \Lambda _{12} \Lambda _{23} \Lambda _{12})\; \sigma ^1 \otimes \sigma ^2 \otimes \sigma ^3 \nonumber \\
\sigma ^1 \tilde \wedge \;(\sigma ^2  \tilde \wedge \; \sigma ^3) &=& ( E - \Lambda _{12} + \Lambda _{12} \Lambda _{23} - \Lambda _{23} + \Lambda _{23}\Lambda _{12} - \Lambda _{23} \Lambda _{12} \Lambda _{23})\; \sigma ^1 \otimes \sigma ^2 \otimes \sigma ^3
\end{eqnarray} 

By requiring the associativity for the new product $\tilde \wedge$, it is very easy to obtain the following constraint:
\begin{equation}
\Lambda _{12} \Lambda _{23} \Lambda _{12} = \Lambda _{23} \Lambda _{12} \Lambda_{23}
\end{equation}

To comment this result we recall that for the algebraic deformation in the context of Fadeev {\it et al}, the main result is related to the Yang-Baxter equation (YBE). The latter plays a fundamental role in the introduction of the quantization of Lie groups and lie algebras. For some special solutions of the YBE, one can obtain interesting deformed lie groups within the classes A,B,C and D. The mathematical meaning of this equation is that it is equivalent to the associativity of the R-algebras of functions on G (G=A,B,C or D).

In our context the frame ground is the same but the purpose here is to build some mathematical structures, through this deformation, leading to the description of statistics generalizing the bosonic and fermionic ones. For this, we focus on the fact that relation (9) is nothing but a representation of the braid group $ B_{n-1} $ generated by $\{ \varepsilon_i / i=1,...n-1 \}$ satisfying:
\begin{eqnarray}
\varepsilon _i \varepsilon_{i+1} \varepsilon _i &=& \varepsilon _{i+1} \varepsilon _i \varepsilon _{i+1} \\
\varepsilon _i \varepsilon {j} &=& \varepsilon _j \varepsilon_i  \;\;\;\;\;\;\;\;\;\; for\;\;\; |i-j|\ge 2 \nonumber
\end{eqnarray}
we point out that these relations are consistent with the equalities (6); all permutations of the type $ i \rightarrow j$ for $ |i-j| \ge 2$ are not considered.

We can recover the YBE by getting $ R=P\Lambda$; $P$ is the permutation matrix. Therefore we obtain:
\begin{equation}
R_{12}R_{13}R_{23} = R_{23} R_{13}R_{12}
\end{equation}

Now, by assuming that the $\Lambda$-deformed wedge product $\tilde \wedge$ is associative, an arbitrary $\Lambda$-deformed three form is written as follows:
\begin{equation}
\omega ^{(3)}_q = \omega _{ijk} \; \sigma ^i \tilde \wedge \sigma^j \tilde \wedge \sigma^k
\end{equation}

the $\omega_{ijk}$'s are complex numbers.

The space $\Omega ^{(3)}_{\Lambda}(V) $ of all $\Lambda$-three forms then is the vector space generated by the set $\{ \sigma ^i \tilde \wedge \sigma^j \tilde \wedge \sigma^k \;\;\; ; i,j,k= 1,...,dim(V) \}$

By induction, we generalize this construction to an arbitrary order $p$. Therefore, an arbitrary element of the space of $\Lambda -p$ forms denoted by $\Omega ^{(p)} _{\Lambda}(V)$ is given by:
\begin{equation}
\omega^{(p)}_{\Lambda} = \omega _{i_1 ... i_p} \sigma^{i_1} \tilde \wedge \sigma^{i_2}\tilde \wedge ... \tilde \wedge \sigma ^{i_p}
\end{equation}
\hspace{4cm} $\omega _{i_1,...,i_p} \epsilon K$

Let us notice that the dimension of $\Omega ^{(p)} _q (V)$ is more than $C^p_r = {r! \over (r-p)! p!} = dim \Omega ^{(p)} _{\Lambda = P}(V)$. This equality is obtained only if the antisymmetry is restored i.e; if and only if we have: $ \Lambda ^{ij}_{kl} = \delta^i _l \delta ^j _k $.

To end this section, we will construct the algebra of $\Lambda$-exterior forms on V. Let us start by considering two different $\Lambda$-forms $\omega ^{(\alpha )}_{\Lambda }$ and $\omega ^{(\beta )}_{\Lambda }$ that belong respectively to $\Omega ^{(\alpha )} _{\Lambda}(V)$ and $\Omega ^{(\beta )} _{\Lambda}(V)$. Their exterior product is given by:

\begin{equation}
\omega ^{(\alpha )} _{\Lambda}(V) \tilde \wedge \; \omega ^{(\beta )} _{\Lambda}(V) = \phi _{i_1,...,i_{\alpha}} \psi _{j_1,...,j_{\beta}} \sigma ^{i_1} \tilde \wedge ... \tilde \wedge \sigma^{i_{\alpha}} \tilde \wedge \sigma ^{j_1} \tilde \wedge ... \tilde \wedge \sigma ^{j_{\beta}}
\end{equation}
with $\omega ^{(\alpha )} _{\Lambda}(V) = \phi _{i_1,...,i_{\alpha}} \sigma ^{i_1}\tilde \wedge ... \tilde \wedge \sigma ^{i_{\alpha }}$ and $ \omega ^{(\beta )} _{\Lambda}(V) = \psi _{j_1,...,j_{\beta}} \sigma ^{j_1}\tilde \wedge ... \tilde \wedge \sigma ^{j_{\beta }}$

The deformed exterior algebra denoted $\Omega _{\Lambda } (V)$is defined in analogy with the classical limit;

\begin{equation}
\Omega _{\Lambda } (V) = \bigoplus ^{\infty} _{ p=0} \Omega ^{(p)}_{\Lambda}(V)\;\;\;,\;\;\; \Omega ^{(0)} _{\Lambda} = K
\end{equation}

Through this result, we point out that this deformed exterior algebra is an infinite dimensional space contrary to the case $\Lambda = P$ where we have:
\begin{equation}
dim_{\Lambda = P} (v) = 2 ^{dim v}
\end{equation}

This is the same result found in \cite{ando1}, where it has been proved using a completely different approach that the space $\Omega _q (V)$ is infinite dimensional.

\section{ Deformation of the exterior algebra $\Omega (M_n)$}

The purpose of this paragraph is to generalize the previous result to the case of a finite manifold $M_n$ instead of the vector space. In order to do this, one have to introduce an adequate differential calculus leading to a consistent deformed wedge product. Let us then start by introducing such a differential calculus. This calculus is based on the differential operator $d$:

\begin{itemize}
\item  i) $d^2 = 0$
\item  ii)$d$ satisfies the graded Leibnitz rule.
\item  iii) The differential calculus is invariant under transformations: $ x^i \rightarrow \alpha _{ij} x^j $ where $x^i$ are the coordinates of a given point on $M_n$.
\end{itemize}

The condition iii) is less restrictive than the invariance $GL_{q_{ij}}$ required in the work \cite{ber} see also \cite{faddeev} (and references their in) for the $SL_q$ case. So, it is of great interest for us to treat with differential calculus defined this way in order to find a convenient deformed wedge product. The realization of this differential calculus in term of partial derivatives of functions on $M_n$ is given by:
\begin{equation}
df(...) = \sum _{i=1} ^n \partial _i f(...) dx^i
\end{equation}
Using these tools, we rewrite eq(3) as:
\begin{equation}
dx^i \tilde \wedge dx^j= dx^i \otimes dx^j - \Lambda ^{ij}_{kl} dx^k \otimes dx^l
\end{equation}

As has been demonstrated in the previous section, the associativity of $\tilde \wedge$ is equivalent to the (YBE) satisfied by $\Lambda$.

To build the algebra of $\Lambda$-deformed exterior forms on $M_n$, $\Omega _{\Lambda}(M_n)$, we have to consider more conditions on $\tilde \wedge$. This is due to the fact that the operator $d$ is  subject to constraints (17). The $\Lambda$-deformed wedge product (18) will be seen as generalizing the ordinary antisymmetric wedge product $\wedge$. We have then to put:
\begin{equation}
dx^i \tilde \wedge dx^j = -S^{ij}_{kl} dx^k \tilde \wedge dx^l
\end{equation}
where $S$ is a $n^2 \times n^2$ matrix, such that $\tilde \wedge \rightarrow \wedge$ when $S\rightarrow P$.

From eq(19), one can obtain by a direct calculation:
\begin{equation}
(E_{12} - S_{12})(E_{12} + \Lambda _{12}) = 0
\end{equation}
One can check that the solution of this equation is given by:
\begin{eqnarray}
\Lambda &=& \sum _i e^i_i\otimes e^i_i + \sum_{i\neq j} q_{ij} e^i_j \otimes e^j_i \\ \nonumber
and & & \\
S &=& \sum _i p_i e^i_i \otimes e^i_i + \sum _{i\neq j} q_{ij} e^i_j \otimes e^j_i \nonumber
\end{eqnarray}

Now, we introduce the deformed $p- forms$ on $M_n$. As for the case of the vector space V, a general deformed $p$-form is written:

\begin{equation}
\omega ^{(p)}_{\Lambda} = \omega _{i_1,...,i_p}(x_1,...,x_n)dx^{i_1}\tilde \wedge ... \tilde \wedge dx^{i_p}
\end{equation}
where the coefficients $\omega _{i_1,...,i_p}$ are functions of the variables $(x_1,...,x_n)$.

Using eq(17), we can construct deformed $(p+1)$-forms from $p$-forms:
\begin{equation}
d\omega ^{(p)}_{\Lambda} = \partial _{i_0} \omega _{i_1,...,i_p}(x_1,...,x_p) dx^{i_0} \tilde \wedge ... \tilde \wedge dx^{i_p}
\end{equation}

The coordinates here are not commuting as in the usual case $(\Lambda = P )$.

In order to have consistency between eqs(19, 20, 21, 23) one have to impose the following commutation relations:

\begin {eqnarray}
x - x & \;\;\;\;\;\;\;\; & x^i. x^j = q_{ij} x^jx^i \nonumber \\
x - dx & & x^i.dx^i = p_i dx^i.x^i \nonumber \\
& & x^i.dx^j = -q_{ij} dx^j.x^i \nonumber \\
dx - dx & & dx^i\tilde \wedge dx^i = 0 \nonumber \\
& & dx^i \tilde \wedge dx^j = -q_{ij}dx^j\tilde \wedge dx^i \\
\partial - x & & \partial _i. x^i = 1+ p_i x^i .\partial _i \nonumber \\
& &\partial _i .x^j= {1 \over q_{ij}}x^j\partial _i \nonumber \\
\partial - dx & & \partial _i.dx^i = {1 \over p_i}dx^i.\partial _i \nonumber \\
& & \partial _i .dx^j = {1 \over q_{ij}} dx^j.\partial _i \nonumber \\
\partial - \partial & & \partial _i . \partial _j = q_{ij} \partial _j \partial _i \nonumber
\end{eqnarray} 

In eqs(24) the product between elements $dx.dx$ is the exterior deformed wedge product. From the algebraic point of view it can be considered as the dot (.) composing the other elements of other relations of (24). In what follows the operation $\tilde \wedge$ and . will be required to commute.

Let us recall that we are now dealing with matrices $\Lambda$ depending on the parameters $q_{ij}$ and $p_i$.

To conclude, we introduce the $\Lambda$-analogue of the Leibnitz rule. We note that the nilpotency condition on $d$ is ensured by the constraint: $q_{ij}q_{ji}=1$ which follows from the first equation in (24).

Concerning the $\Lambda$-analogue of Leibnitz rule, it must generalize the usual one;
\begin{equation}
d(\psi \wedge \eta) = d\psi\wedge \eta + (-1)^{deg \psi} \psi \wedge d\eta
\end{equation}
with $\psi$ and $\eta$ arbitrary forms in $\Omega (M_n)$

In the case of the $\Lambda$-deformed wedge product, a generalization of this expression is not obvious, this is the reason why we look for a $\Lambda$-analogue of it in the spirit of what was done in the work \cite{wess}. Indeed, the authors consider functions $f(x_1,...,x_n)$ at a given point of $M_n$ as polynomials on the coordinates. This form of the functions allows us to define the Leibnitz rule corresponding to the deformed wedge product. Let us thus start by discussing the case of two $\Lambda$-deformed 1-forms, $\psi$ and $\eta$ given by:
\begin{eqnarray}
\psi &=& \psi _{i_1}(x_1,...,x_n)dx^{i_1} \nonumber \\
\eta &=& \eta _{j_1}(x_1,...,x_n)dx^{j_1}
\end{eqnarray}

The $\Lambda$-Leibnitz rule corresponding to these forms is obtained by computing the following:
\begin{equation}
d(\psi \tilde \wedge \eta) = d(\psi _{i_1}(x_1,...,x_n) dx^{i_1} \tilde \wedge \eta _{j_1} (x_1,...,x_n) dx^{j_1})
\end{equation}
we take:
\begin{eqnarray}
\psi _{i_1} (x_1,...,x_n) &=& \sum _{\mu_1,...,\mu_n} a_{\mu_1...\mu_n}x_1^{\mu_1}...x_n^{\mu_n}\nonumber \\
\eta _{j_1}(x_1,...,x_n) &=& \sum _{\nu_1,...,\nu_n} b_{\nu_1...\nu_n}x_1^{\nu_1}...x_n^{\nu_n}
\end{eqnarray}
A direct calculation using eq(21) leads to:
\begin{eqnarray}
\partial_{i_0}(x_1^{\mu_1}...x_n^{\mu_n}) = & & \Big( \prod _{k=1}^{i_0-1} q_{i_0k}^{-\mu_k}\Big) {[\mu_{i_0}]}_{p_{i_0}}\; x_1^{\mu_1}...x_i^{\mu_i-1} ...x_n^{\mu_n} \nonumber \\
& & + \prod ^n _{k=0 \; k\neq i_0} q_{i_0k}^{-\mu_k} p_i^{\mu_i} \; x_1^{\mu_1}...x_{i_0}^{\mu_{i_0}}...x_n^{\mu_{n}}\partial_{i_0}
\end{eqnarray}
with 
\[
[x]_q = {1 - q^x \over 1 - q}
\]
One can then get:

\begin{eqnarray}
d(\psi \tilde \wedge \eta ) &=& d\Big\{ \sum _{(\nu)(\mu)} a_{\mu_1...\mu_n}b_{\nu_1...\nu_n} x_1^{\mu_1}...x_n^{\mu_n}dx^{i_1}\tilde \wedge x_1^{\nu_1}...x_n^{\nu_n}dx^{j_1} \Big\} \nonumber \\
&=& d \Big\{ \sum _{(\nu)(\mu)} a_{(\mu)}b_{(\nu)} \prod ^n _{l=1 \; l\neq i_1} {(-{q_{l i_1}})}^{-\nu_l} p_{i_1}^{-\nu_{i_1}} \; (x_1^{\mu_1}...x_n^{\mu_n}x_1^{\nu_1}...x_n^{\nu_n})dx^{i_1}\tilde \wedge dx^{j_1} \Big\}  \\
&=& \sum _{(\nu)(\mu)} \sum_{i_0} a_{(\mu)}b_{(\nu)} \prod ^n _{l=1 \; l\neq i_1} {(-{q_{l i_1}})}^{-\nu_l} p_{i_1}^{-\nu_{i_1}} \partial_{i_0} (x_1^{\mu_1}...x_n^{\mu_n}x_1^{\nu_1}...x_n^{\nu_n}) \; dx^{i_0} \tilde \wedge dx^{i_1}\tilde \wedge dx^{j_1} \nonumber
\end{eqnarray}
by computing the partial derivative in the last line using (29), we finally get:
\begin{equation}
d(\psi \tilde \wedge \eta ) = \delta _1 (\psi )\; \tilde \wedge \; \eta  + (-1)^{deg \psi } \psi \; \tilde \wedge \; \delta_2(\eta )
\end{equation}
where
\begin{equation}
\delta _1(\psi)= \sum_{\mu} \sum _{i_0} a_{(\mu)} \Big( \prod ^n _{l=1\; l\neq i_0} -q^{\nu_{l}}_{l \, i_0}\Big) \Big( \prod ^{i_0-1}_{k=1} -q^{-\mu_{k}}_{i_0\, k}\Big) p_{i_0}^{\nu_{i_0}} [\mu_{i_0}]_{p_{i_0}}\; x_1^{\mu_1}...x_i^{\mu_{i-1}}...x_n^{\mu_n}. dx^{i_0} \tilde \wedge dx^{i_1}
\end{equation}

\[
\delta _2(\eta)= \sum_{\nu} \sum _{i_0} b_{(\nu)} \Big( \prod ^n _{l=1\; l\neq i_0} -q^{\mu_{l}}_{l \, i_0}\Big) \Big( \prod ^{i_0-1}_{k=1} -q^{-\nu_{k}}_{i_0\, k}\Big) p_{i_0}^{\mu_{i_0}} [\nu_{i_0}]_{p_{i_0}} ( -q_{i_1 i_0}) \; x_1^{\nu_1}...x_i^{\nu_{i-1}}...x_n^{\nu_n}.  dx^{i_0} \tilde \wedge dx^{j_1}
\]

The aspect of the above expression is not very elegant owing to the complication of the non-commuting differential calculus used here. Nevertheless one can recover the usual Leibnitz rule by taking $q_{ij}$ and $p_i$ $=1$, the operators $\delta _1$ ,$\delta _2$ and $d$ are the same in this case. The generalization to two arbitrary forms will be obvious if we proceed in the same way.

\section{generalized statistics}

Using the above tools concerning the $\Lambda$-deformed exterior algebra, we are now in a position to construct a consistent statistical theory. These statistics are neither bosonic nor fermionic but some kind of a generalization of both. We will show that this statistics involve an arbitrary phase, in the algebraic commutation relations instead of the minus or plus sign for fermions and bosons respectively. Returning to the (YBE), we will show that this equation is behind all the construction, starting from the remark that it is nothing but a mathematical representation of a braid relation appearing in intermediate statistics. We will also opportunate on how to construct the multiparticle wave functions.

Let us first recall how to construct the usual Fock space representation (fermionic and bosonic one). We denote by $H^n$, the n-particle space that is nothing but the n-tensor product of a given Hilbert space $H$. The Fock space over $H$ is defined by:
\begin{equation}
F(H) = \bigoplus _{n\ge 0} H^n
\end{equation}

The elements of $F(H)$ are given by the set:
\[
\{ \varphi \epsilon F(H) : \varphi = (\varphi^{(0)}, \varphi^{(1)},...,\varphi^{(n)},...)\;\; ; \, \varphi^{(i)} \epsilon H^i \}
\]

The finite particle subspace $F^0(H) \subset F(H)$ is defined by: $\varphi \epsilon F^0(H)$ if and only if $\varphi^{(n)} = 0$ for $n\ge 1$.

Now we define the annihilation and creation operators on $ F (H)$ starting from the action of the operators $P^- (f)$ and $ P^+(f)$ on the set:
\begin{equation}
D^n = \{f_1 \otimes f_2 \otimes ... \otimes f_n : f_i \epsilon H \} \subset H^n ; D^0 = C^{(1)}
\end{equation}
as
\begin{eqnarray}
P^-(f) &:& D^n \longrightarrow D^{n-1}\;\;\;\;\;\; n\ge 1 \\
P^+(f) &:& D^n \longrightarrow D^{n+1}\;\;\;\;\;\; n\ge 0 \nonumber
\end{eqnarray}
and 
\begin{eqnarray}
P^-(f)(f_1\otimes ... \otimes f_n) &=& (f,f_1)(f_2\otimes ... \otimes f_n) \\
P^+(f)(f_1\otimes ... \otimes f_n) &=& (f \otimes f_1\otimes ... \otimes f_n)  \;\; \forall \,f \epsilon H \nonumber
\end{eqnarray}

For $n=0$ one sets:
\begin{equation}
P^-(f) H^0 = 0
\end{equation}

The bilinear form (.,.) in eq (36) is the scalar product on the considered Hilbert space $H$.

The description of bosons and fermions is introduced through the subsets $H^n _{\pm}$ of $H^n$. They are respectively the totally symmetric and antisymmetric n-fold tensor of H.

The corresponding bosonic and fermionic Fock spaces are:
\begin{equation}
F_{\pm}(H) = \bigoplus _{n\ge 0} \;\;\;H_{\pm}^n
\end{equation}
To construct the bosonic and fermionic creation and annihilation operators one introduces the operators:
\begin{eqnarray}
B^{\pm}(f) &=& A P^{\pm}(f) A \\
C^{\pm}(f) &=& S P^{\pm}(f) S \nonumber
\end{eqnarray}
where $A$ and $S$ are the antisymmetric and the symmetric projection operators on $F(H)$ respectively:
\begin{eqnarray}
A(f_i \otimes f_j) &\equiv & f_i \wedge f_j = {1\over 2}(f_i \otimes f_j - f_j \otimes f_i) \\
S (f_i \otimes f_j) &\equiv & f_i \vee f_j = {1 \over 2} (f_i \otimes f_j + f_j \otimes f_i) \nonumber 
\end{eqnarray}
With these equalities we introduce the sets:
\[
D^n _+ \equiv  \{ f_1 \vee f_2 \vee ... \vee f_n ; \;\;f_i \epsilon H\}
\]
and
\begin{equation}
D_-^n \equiv  \{ f_1 \wedge f_2 \wedge ... \wedge f_n ; \;\; f_i \epsilon H\} 
\end{equation}

The $H^n_{\pm}$'s are the linear combinations of the elements of $D^n _{\pm}$, respectively.

This is how things are done in the classical case (non-deformed case $ \rightarrow $ Fermions and Bosons). In order to obtain generalized statistics, we will need the previously defined wedge product. In fact, to generalize the equations above, one considers the following $\Lambda$-deformed exterior product:

\begin{eqnarray}
f_i \; \tilde \wedge \; f_j &=& {1\over 2} ( f_i \otimes f_j + \Lambda _{ij}^{kl} f_k \otimes f_l) \\
&\equiv & \Omega (f_i \otimes f_j) \nonumber
\end{eqnarray}

This equation unifies the two relations (40). Indeed, by taking $\Lambda = -P$ or $P$ one recover classical statistics.

Now, we are going to construct the oscillator algebra corresponding to this exterior product. To start we rewrite relation (42) in a condensed form as:
\begin{eqnarray}
f_1 \; \tilde \wedge \; f_2 &=& {1\over 2}(f_1 \otimes f_2 + \Lambda_{12} f_1 \otimes f_2 ) \\
&\equiv & \Omega _{(2)} (f_1 \otimes f_2) \nonumber
\end{eqnarray}
with $\Omega_{(2)} = {1 \over 2}(E_{12} + \Lambda _{12})$

As we have seen before, the associativity of $\tilde \wedge$ is equivalent to the Braid equation on $\Lambda$.

The generalized statistics using $\Omega$ instead of $A$ or $S$ above are obtained by introducing the following space:
\begin{equation}
\tilde F(H) = \bigoplus _{n\ge 0} \tilde H^n
\end{equation}
where $\tilde H^n$ is the set of elements:
\[
\tilde \varphi ^{(n)} = \varphi^{i_1,...,i_n}f_{i_1} \tilde \wedge ... \tilde \wedge f_{i_n}
\]
$\varphi^{i_1, ... ,i_n}$  are just complex numbers and $f_1 \tilde\wedge ... \tilde\wedge f_n$ generate the set $\tilde D^n$.

The Hilbert space can be considered as $L^2(R^s, d^sx)$. We introduce thus the operators $\{ \varepsilon _i : i=1,...,p-1 \}$ acting on $ \tilde D^p$ as:

\begin{eqnarray}
(\varepsilon_i \tilde \varphi^{(p)})(x_1,...,x_p) &=& [\Lambda_{i i+1}\varphi ^{(p)}](x_1,...,x_p) \nonumber \\
&\equiv & \left\{ 
\begin{array}{cc}
q(x_i,x_{i+1}) & \varphi ^{(p)} (x_1,...,x_{i+1},x_i,..,x_p) \\
0  & i\ge p 
\end{array}
\right.
\end{eqnarray}
where $\varphi^{(p)} \equiv f_1 \otimes ... \otimes f_p \epsilon D^p ; x_i \epsilon R^s$ and $q(x_i,x_{i+1})$ is an exchange factor.

At this level, we notice that the matrices in (45) are particular in the sense that

\begin{itemize}
\item i) they depend on coordinates $x_i$
\item ii) they have the special form; $\Lambda _{il}^{kl}(x_i,x{i+1}) = \delta _{ip} \delta _{jk} q(x_i,x_{i+1})$
\end{itemize}

Moreover, the objects $\varepsilon$ appearing in (45) and defined through $\Lambda$, are subjects to the following constraints:
\begin{eqnarray}
\varepsilon_i \varepsilon_{i+1} \varepsilon_i &=& \varepsilon_{i+1} \varepsilon_i \varepsilon_{i+1} \nonumber \\
\varepsilon_i \varepsilon_j &=& \varepsilon_j \varepsilon_i \;\;\;\;\;\; for \;\; |i-j| \ge 2 
\end{eqnarray}

Notice that the matrices $\Lambda$ are introduced in this way to be a mathematical representation of the Braid relations on $\varepsilon_i$.

Consequently, a general element of $\tilde D^3$ has the generic form:
\begin{eqnarray}
f_1\tilde\wedge f_2 \tilde\wedge f_3 &=& {1\over 6}(E - \Lambda_{12} - \Lambda_{23} + \Lambda_{12}\Lambda_{23} + \Lambda_{23}\Lambda_{12} + \Lambda_{12}\Lambda_{23}\Lambda_{12})f_1\otimes f_2 \otimes f_3 \\
& \equiv & \Omega_{(3)} f_1 \otimes f_2 \otimes f_3 \nonumber
\end{eqnarray}

In the same way we generalize this construction from 3 to $n$ as:
\begin{equation}
f_1 \tilde\wedge f_2 \tilde\wedge ... \tilde\wedge f_n \equiv \Omega _{(n)} f_1 \otimes f_2 \otimes ... \otimes f_n
\end{equation}

In what follows, we introduce the creation and annihilation operators. These are constructed in analogy with the classical case, relations (39). We define the elements $B^+(f)$ and $B^-(f)$ as:

\begin{equation}
B^{\pm}(f) = \Omega _{(n\pm1)} P^{\pm}(f) \Omega _{(n)}, \;\;\; f\epsilon H
\end{equation}

Their actions on $\tilde D^n$are given by:
\begin{eqnarray}
(B^+(f) \tilde\varphi^{(n)})(x_1,...,x_n) &\equiv & \int d^sx \bar f(x) \varphi ^{(n+1)} (x,x_1,...,x_n) \nonumber \\
(B^-(f) \tilde\varphi^{(n)})(x_1,...,x_n) &\equiv & {1\over n} f(x_1)\varphi^{(n-1)}(x_2,...,x_n) \nonumber \\
&+& {1\over n} \sum _{k=2}^n q(x_1,x_k)...q(x_{k-1},x_k)f(x_k)\nonumber \\
& \times & \varphi ^{(n-1)} (x_1,...,\hat x_k,...,x_n)
\end{eqnarray}
The notation $\hat x_k$ in this formula mean that the coordinate is omitted.

To find the commutation relations between these operators we introduce the operators $A^+(x)$ and $A^-(x)$:
\begin{eqnarray}
B^+(f) &=& \int d^sx \bar f(x) A^-(x) \\
B^- (f) &=& \int d^sx \bar f(x) A^+(x)\nonumber
\end{eqnarray}

by a direct calculation
\[
\bigg( A^+(x) \varphi^{(n)}\bigg) (x_1,...,x_n) = \varphi^{(n+1)} (x,x_1,...,x_n)
\]
and
\begin{eqnarray}
\bigg( A^-(x)\varphi^{(n)}\bigg)(x_1,...,x_n) = &&{1\over n} \delta(x-x_1)\varphi^{(n-1)}(x_2,...,x_n) \nonumber \\
&+& {1\over n} \sum^n_{k=2} q(x_1,x_k)...q(x_{k-1},x_k) \delta(x-x_k) \nonumber \\
&\times & \varphi^{(n-1)}(x_1,...,\hat x_k,...,x_n) 
\end{eqnarray}

\begin{eqnarray}
(n+1)A^-(x_1)A^+(x_2) - q(x_1,x_2)nA^+(x_2)A^-(x_1) &=& \delta(x_1-x_2) \nonumber \\
A^-(x_1)A^-(x_2) - q(x_2,x_1) A^-(x_2)A^-(x_1) &=& 0 \\
A^+(x_1)A^+(x_2) - q(x_2,x_1) A^+(x_2)A^+(x_1) &=& 0 \nonumber
\end{eqnarray}
These relations hold when the parameters $q$ satisfy:

\begin{eqnarray}
q(x_i,x_{i+1})q(x_{i+1},x_i) = 1 \\
q(x_i,x_{i+1}) = \bar q(x_{i+1},x_i) \nonumber
\end{eqnarray}

Now consider the operator denoted by $f(N)$ and acting on a given element $ \phi = (\xi_0, \xi_1,...,\xi_n,...) \;\;\;;\;\;\;\xi_i \epsilon \tilde D^i \;\;(i=0,...,n)$ as:
\begin{equation}
f(N)\phi = (f(0)\xi_0,f(1)\xi_1,...,f(n)\xi_n,...)
\end{equation}

$f(N)$ is an operator in the Fock space. It is used to rewrite the relations (53). Let us first note that it satisfies the following properties:
\begin{eqnarray}
f(N)\Omega_{(n)} &=& \Omega_{(n)} f(N) \nonumber \\
f(N) P^+(f) &=& P^+(f) f(N+1) \nonumber \\
f(N) B^+(f) &=& B^+(f) f(N+1) \\
f(N+1) P^-(f) &=& P^-(f) f(N) \nonumber \\
f(N+1) B^-(f) &=& B^-(f) f(N) \nonumber
\end{eqnarray}
with the following change:
\begin{eqnarray}
a^+(x) = {\sqrt N}A^+(x) &=& A^+(x) \sqrt {N+1} \\
a^-(x) = A^-(x) {\sqrt N} &=& \sqrt {N+1} A^-(x) \nonumber
\end{eqnarray}
(53) becomes:
\begin{eqnarray}
a^-(x_1)a^+(x_2) - q(x_1,x_2)a^+(x_2)a^-(x_1) &=& \delta(x_1-x_2)\nonumber \\
a^-(x_1)a^-(x_2) - q(x_2,x_1)a^-(x_2)a^-(x_1) &=& 0 \\
a^+(x_1)a^+(x_2) - q(x_2,x_1)a^+(x_2)a^+(x_1) &=& 0 \nonumber
\end{eqnarray}

By considering this, we notice that we introduced here the framework for the generalization of the ordinary statistics. In this context we find for a particular case how to extract the commutation relations (58) that coincides with the ones corresponding to anyonic oscillators. Moreover one can recover the relation:
\begin{equation}
(a^-(x))^2 = (a^+(x))^2 = 0
\end{equation}
by taking $x_1=x_2=x$. this equality is nothing but the equation ensuring the Pauli exclusion principle for Anyons.

At this step of work it is important to notice that through this process of deformation we can build up also the multiparticle function allowing the description of a system of particles. This is done in a general case, one have to consider just the definition of the exterior deformed product introduced in the beginning of this work. By taking some special solutions of the (YBE) it is possible to obtain all intermediate statistics.

We conclude this part of the work by giving the wave functions corresponding to two particles:

\begin{equation}
\tilde \psi (x_1,x_2) = {1\over 2}(f_1(x_1) f_2(x_2) + q(x_1,x_2)f_2(x_1) f_1(x_2))
\end{equation}
This study can be generalized to $n$ particles as follows:
\begin{equation}
\tilde \psi (x_1,...,x_n) = \bigg(\Omega _{(n)} (f_1 \otimes f_2 \otimes ... \otimes f_n)\bigg)(x_1,...,x_n)
\end{equation}
The exchange of the position of two particles leads to a phase factor that depends on the position of these two particles. This behaviour is proper to anyons \cite{leinaas,lerda}

\section{Application to quons; Intermediate statistics}
\subsection*{quons:}

Here, we will deal with the tools introduced above concerning this deformation scheme; the deformation of the exterior wedge product. We introduce in this context the particles called quons in the literature of intermediate statistics. these are obtained by considering a particular solution of the (YBE), this is done by choosing the particular deformed wedge product:
\begin{equation}
f_i \tilde\wedge f_j = {1\over \sqrt 2}(f_i\otimes f_j + q f_j\otimes f_i)
\end{equation}
The basic relations on which our approach is based are rewritten now:
\begin{eqnarray}
(f_i\otimes f_j)\tilde \wedge f_k &=& {1\over \sqrt 3}(f_i\otimes f_j\otimes f_k + qf_i\otimes f_k\otimes f_j + q^2 f_k\otimes f_i\otimes f_j) \\
f_i \tilde \wedge (f_j\otimes f_k) &=& {1\over \sqrt 3}(f_i\otimes f_j\otimes f_k + qf_j\otimes f_i\otimes f_k + q^2 f_j\otimes f_k\otimes f_i)\nonumber
\end{eqnarray}
In a compact form, we rewrite them as:
\begin{eqnarray}
f_1\tilde \wedge (f_2\otimes f_3) &=& {1\over \sqrt3}( E + qP_{12} + q^2 P_{12}P_{23})f_1\otimes f_2\otimes f_3 \\
(f_1\otimes f_2) \tilde \wedge f_3 &=& {1\over \sqrt 3} (E + qP_{23} + q^2 P_{23}P_{12})f_1\otimes f_2\otimes f_3\nonumber
\end{eqnarray}

So, in this particular solution of the (YBE), we obtain the three deformed form as:
\[
(f_1\tilde \wedge f_2)\tilde \wedge f_3 = Q_{(3)} (f_1 \otimes f_2\otimes f_3)
\]
with 
\begin{equation}
Q_{(3)} = (E + qP_{12} +qP_{23} +q^2 P_{12}P_{23} + q^2 P_{23}P_{12} + q^3 P_{23} P_{12} P_{23} )
\end{equation}

The (YBE) is satisfied when we require the product $\tilde\wedge$ to be associative.

Using the tools discussed above we construct the generalized statistics. We reproduce the expression of an n-deformed form as:
\begin{eqnarray}
\psi(x_1,x_2,...,x_n) &\equiv & (f_1 \tilde\wedge ... \tilde\wedge f_n) (x_1,...,x_n) \\
&=& Q_n(f_1\otimes ... \otimes f_n)(x_1,...,x_n) \nonumber
\end{eqnarray}

with $\displaystyle Q_n = {1\over \sqrt n} \sum _{P\epsilon S_n} q^{m(P)} P$, $S_n$ is the permutation group and $m(P)$is the minimal number of transpositions generated by the permutation $P$.

It is obvious to see that $Q_n$ reduces to the symmetric operator $S$ and the anti-symmetric one $A$ by taking respectively $q=1$or $q=-1$.

The main idea in this section is the introduction of the creation and annihilation operators $a^+$ and $a$ defined in the following:
\begin{eqnarray}
a^+ : D^n_q &\longrightarrow & D^{n+1}_q \nonumber \\
\underbrace{f\tilde\wedge f\tilde\wedge ... \tilde\wedge}_{n factors} & \longmapsto &\underbrace{f\tilde\wedge f\tilde\wedge ... \tilde\wedge}_{n+1 factors} \nonumber \\
and \\
a : D^n_q &\longrightarrow & D^{n-1}_q \nonumber \\
\underbrace{f\tilde\wedge f\tilde\wedge ... \tilde\wedge}_{n factors} & \longmapsto & [n]_q \underbrace{f\tilde\wedge f\tilde\wedge ... \tilde\wedge}_{n-1 factors} \nonumber 
\end{eqnarray}
and $a D^0 = 0$

Using these equations, we easily check through (58) that $a$ and $a^+$ obey:
\begin{eqnarray}
[ a , N] &=& a \nonumber \\
aa^+ - qa^+a &=& 1 \\
\hbox{[} a^+ , N ] &=& - a^+ \nonumber
\end{eqnarray}
This is an important result that we recover, through this approach of the q-deformed wedge product: the algebra of quons. We obtain the k-fermionic algebra from this latter by taking $q= \exp{2\pi i {1\over k}}$ and permits the description of the intermediate statistics that interpolate between fermionic ($k=2$) and bosonic ($k=\infty $) ones.

We notice that following this approach we obtain also relations between operators of the same kind:
\begin{eqnarray}
aa - q^{-1}aa &=&0 \nonumber \\
a^+a^+ - q^{-1}a^+a^+ &=& 0 \nonumber
\end{eqnarray}
So these relations arise in a natural way, instead of being imposed to obtain the so-called generalized quons \cite{marcinek}.

Now, and by referring to the classical limit ($ q=1 $or $q=-1$), we will give the statistical properties of the quons Algebra (68).

\subsection*{Partition function and quantum distribution}
In this part we give the partition function corresponding to an ideal $q$-gas. Let us first define the Hamiltonian as:
\begin{equation}
H = \sum_{\lambda} H_{\lambda}
\end{equation}
where $\displaystyle H_{\lambda} = (E_{\lambda} - \mu ) N_{\lambda}$.

$\mu$ is the chemical potential, $E_{\lambda}$ and $N_{\lambda}$ are respectively the kinetic energy and number operator for quons in a given $\lambda$-mode.

The q-analogue of the base factor for a $\lambda$-mode is:
\begin{equation}
(f_{\lambda})_q = {1 \over Z} Tr(e^{-\beta H}a^+_{\lambda}a_{\lambda})
\end{equation}
$\displaystyle Z = Tr e^{-\beta H}$ is the partition function and $\beta = (K_BT)^{-1}$ the reciprocal temperature ($K_B$ is the Boltzmann constant). We notice here that it is important to distinguish between the two (physically different) cases $|q|\neq 1$ and $|q| = 1$. In fact while in the former the Fock space is infinite dimensional, it is finite in the latter.
One can verify that for a complex $q$, the partition function:
\begin{equation}
Z = \prod _{\lambda} {1\over 1-e^{-\eta}} \;\;\;\; , \;\;\;\; \eta = \beta (E_{\lambda} -\mu)
\end{equation}
is not $q$-dependent. The base factor $(f_{\lambda})_q$ reads as:
\begin{equation}
(f_{\lambda})_q = (1-e^{-\eta}) Tr(e^{\eta N_{\lambda}}[N_{\lambda}])
\end{equation}
It converges for $|q| < e^{-\beta \mu}$. As a result, we obtain
\begin{equation}
(f_{\lambda})_q = {1\over e^{\eta} -q}
\end{equation}

In the classical limit $q=1$ or $q=-1$, one recovers the ordinary Bose and Fermi distributions. Between these two values can locate intermediate statistics described by $q$ having the expression $ q=exp{2\pi i\over k}$. For this value, the partition function can be computed Via the generalized Pauli exclusion principle. We have then:
\begin{equation}
Z= \prod_{\lambda} Z_{\lambda} \;\;\;\;\;\; Z_{\lambda} = \sum ^{k-1}_{n_{\lambda} = 0}e^{-\eta n_{\lambda}} = { 1- e^{-\eta k} \over 1-e^{-\eta}}
\end{equation}
This formulae can be seen  as an unification of the Fermi and Bose partition function respectively for $k=2$ and $k=\infty$. 

Now, we are able to derive the occupation number
\begin{equation}
(f_{\lambda})_q = {1\over Z} \sum ^{k-1}_{n_{\lambda}=0}[n_{\lambda}]e^{-\eta n_{\lambda}}
\end{equation}
with $\displaystyle [n_{\lambda}] = { q^{n_{\lambda}} - 1 \over q-1}$
This equation leads to
\begin{equation}
(f_{\lambda})_q = {1\over e^{\eta} - q}
\end{equation}

Here also we mean by this equality, the distribution of the quantum gas satisfying the commutation relation (68) corresponding to quons. In fact, by taking $k=2$ and $k \rightarrow \infty$, one recovers respectively the Fermi and Bose distribution. It clearly shows that for a complex $q$ or $q=e^{2\pi i\over k}$ our q-gas is described by the same quantum distribution $(f_{\lambda})_q$ playing in some sense a fundamental role in the derivation of the thermodynamical functions. In an elegant form $(f_{\lambda})_q$ could be written as an integer series as:
\begin{equation}
(f_{\lambda})_q = \sum_{j\ge 0} e^{-\eta(j+1)q^j}
\end{equation}
it reduces ($k=\infty ; q=1$ and $k=2 ; q=-1$) to the series $\sum e^{-\eta j}$ for bosons and $\sum e^{-\eta j}(-1)^{j+1}$ for fermions.

We can also discuss the Bose-Einstein condensation corresponding to quons constituting our physical system. We enclose them in a large D-dimensional volume $V(D)$. The distribution $(f_{\lambda})_q$ becomes a function depending on a continuous parameter $\epsilon = \gamma^{-1} p^{\alpha}$ ($\alpha = 1,2$), where $\alpha =1$ or 2 and corresponds respectively to the ultra-relativistic or non-relativistic q-gas. $\gamma$ is then 1 or $2n$. It is easy to treat, basing on these tools the generalized Sommerfeld expansion of these special particles.

\section{Coherent States}

Now, we turn to the problem of constructing Coherent states ({\bf CS}) related to these deformed oscillators. Following Klauder \cite{klauder63, klauder85}, the definition of CS is essentially based on three properties:
\begin{itemize}
\item i. {\it normalisability}
\item ii. {\it continuity}
\item iii. resolution of unity
\end{itemize}

In the case of the bosonic Harmonic oscillator ({\bf HO}) these states are constructed {\it via} the Fock states as follows:
\begin{eqnarray}
|z> &=& \exp(-{|z|^2 \over 2}) \exp(za^+) |0> \nonumber \\
 &=& \exp(-{|z|^2 \over 2}) \sum _{n \ge 0} {z^n \over {\sqrt{n!}}}|n>
\end{eqnarray}    
where $z$ is a complex number and $|n>$ is the usual Fock space orthonormal basis,
\[
|n> = {(a^+)^n\over \sqrt{n!}} |0>\; , \; \; \;\; a|0> = 0
\]

These states are eigenvectors of the annihilation operator ($a|z> = z|z>$), normalized, continue in the label $z$ and provide a resolution of unity:
\begin{equation}
\int {d^2z\over \pi} |z><z| = 1 \;\;\;\; \;\;\;\;\;\;\; d^2z = dRe(z)dIm(z)
\end{equation}

For the deformed bosons, it is still not clear how such states should be constructed. In fact, in this literature, we distinguish two main approaches concerning the fulfilment of condition iii. i.e. resolution of unity:

\begin{itemize}
\item In the first approach we deform the concept of integration and differentiation in such a manner
that these satisfy similar properties of those obeyed by ordinary ones: $\displaystyle \int {d_q^2z \over \pi}|z><z| = I$ This was done in \cite{arik, per96} for the particular case $ aa^+ - qa^+a =1$, this corresponds to the conventional quons . 

\item while in the other approach the concepts of integration and differentiation are intact (not deformed), and try
to fulfill (b) in an analogous manner as in (79) i.e. find the appropriate weight function (in (79) this weight function is $1\over \pi$, such that (b) holds \cite{pen1, pen2, pen3, kps}
\end{itemize}
Neither of these two approaches is privileged in the literature. To summarize this we consider two examples of deformed bosons.

These two examples giving an idea on the possible ways to generalize the notion of CS from bosons  to deformed bosons; we discuss another example: CS for the $Z_3$-graded HO, this one generalizing the case of fermions (which, as we're going to see, presents drastically differences from the bosonic case).

\subsection{Deformed bosons}

Many deformations of the usual bosonic HO appeared during the last decades. Of course all these cases can be derived directly from the general deformed exterior algebra we presented in the previous sections.

On a purely algebraic background, we have proposed in \cite{mine1} an algebra ${\cal A}_q$ that generalizes all these deformations. The algebra ${\cal A}_q$ is generated by the triplet $\{a, a^+, I \}$, and is
defined through the following "q-mutation" relations:

\[ 
[a,a^+]_q  =  aa^+ - qa^+a = \Delta'_q  
\] 
\begin{equation} 
[a,\Delta _q]_q  =  a\Delta _q - q\Delta _qa = \Delta'_qa  
\end{equation} 
\[ 
[\Delta _q, a^+]_q = \Delta _q a^+  - qa^+\Delta_q = a^+\Delta' _q
\] 
\hspace{6cm}$ \vdots  $ 

\noindent where $q$ is a complex parameter (of deformation), $\Delta _q =
a^+a$, and $ \Delta _q'$ is to be interpreted as a "q-derivative" of $\Delta
_q$.

The Fock space basis associated is defined in the usual way: Given a vacuum
state $|0>$; ($a|0> = 0$) the different number states are generated through the
action of the creation operator on this state:
\begin{equation}
(a^+)^n |0> = \sqrt{[n]_q!} |n>
\end{equation}
where the factorial function is defined as:
\[
[n]_q! = [n]_q [n-1]_q....[1]_q
\]
\[
[0]_q!=1
\]
and the function $[n]_q$ appears in:
\begin{eqnarray} 
a|n> & = & \sqrt{[n]_q} \;|n-1> \nonumber \\ 
a^+ |n> & = & \sqrt{[n+1]_q} \; |n+1> \nonumber \\ 
\Delta_q |n> & = & [n]_q \; |n> \\ 
\Delta'_q |n> & = & ([n+1]_q -q[n]_q) \; |n> \nonumber 
\end{eqnarray} 

A general method to construct CS for the algebra ${\cal A}_q$ in its general form is still missing.\cite{mine1, mine2}

Let's now discuss in detail the construction of CS for two particular cases of ${\cal A}_q$.

\subsection*{$\displaystyle  [n]_q = { 1-q^n \over 1-q} \;\;\; ; \;\;\; q\epsilon R \;\;\; ; \;\;\; 0\le q \le 1$}

\vspace{1cm}
We carry out the construction in this case following \cite{per96, arik} using the first of the two approaches mentioned in the beginning of this section.

For this $[n]_q$ we have the following relations:

\begin{equation}
aa^+ - qa^+a = I \;\;\;\;\; [a,I] = [a^+,I]=0
\end{equation}

The system of CS is obtained, in analogy with the non-deformed case, starting from the Fock states:
\begin{eqnarray}
||q,z> &=& exp_q (za^+) |0> \\
&=& \sum ^{\infty} _{n=0} {z^n \over \sqrt{[n]_q!}}|n> \nonumber
\end{eqnarray}
where we introduced the deformed exponential function $\displaystyle exp_q(x) = \sum ^{\infty}_{n=0} {x^n \over [n]_q!}$, which converge for $ |x| < R_q = (1-q)^{-1}$ and satisfies:
\begin{equation}
\bigg( {d\over dx} \bigg)_q exp_q(x) = exp_q(x)
\end{equation}
where the q-derivative is defined by:
\begin{equation}
\bigg({d\over dx}\bigg)_q f(x) = {f(x) -f(qx) \over x(1-q)}
\end{equation}
we can also show that:
\begin{equation}
exp_q(x) = {1\over \prod _{k=0}^{\infty}\big( 1-q^k(1-q)x\big)}
\end{equation}
Let's now, check that the states (84) are indeed CS:

\begin{itemize}
\item (84) are eigenstates of the annihilation operator: $a||q,z> = z ||q,z>$

\item The states are clearly normalizable:
\begin{equation}
<q,z||q,z> = \sum _{n=0} {|z|^{2n} \over [n]_q!} = exp_q(|z|^2)
\end{equation}
So the normalized states are given by:
\begin{eqnarray}
|q,z> &=& {\cal N}(|z|^2)||q,z> \\
&=& \bigg( exp_q(|z|^2)\bigg)^{-{1\over 2}} \sum _{n=0} {z^n \over \sqrt{[n]_q!}}|n> \nonumber
\end{eqnarray}
These states are not orthogonal:
\begin{equation}
<q,z|q,z> = {\cal N}(|z|^2) {\cal N}(|z'|^2) exp_q(\bar z z')
\end{equation}

\item The continuity of the states is ensured by the continuity of the overlap term (90) through:
\begin{equation}
\bigg| |q,z> - |q,z'> \bigg| ^2 = 2 \bigg( 1 - Re<q,z|q,z'> \bigg)
\end{equation}

\item A resolution of unity in terms of these states is also possible through the indefinite integral, the so-called Jackson integral \cite{gasper} 
\begin{equation}
\int_0^a f(x) d_qx = a(1-q) \sum_{n=0}^{\infty} q^kf(q^ka)
\end{equation}
which satisfies $\displaystyle \int _0^{x_1} (exp_q(x))^{-1} x^n d_qx = [n]_q! $

The resolution of unity is giving by:
\begin{equation}
{1\over \pi} \int_{D_q}d^2_q z |z><z| = {1\over 2\pi} \int _0^{2\pi} d\theta \int _0^{R_q^2} d_q(r^2) |z><z| = 1
\end{equation}
$\displaystyle z=r e^{i\theta} \;\;\;,\;\;\; D_q = \{z : |z| < R_q\}$

\end{itemize}
This completes the construction of CS using the first approach for this case. To discuss the second approach we take another example:

\subsection*{$\displaystyle [n] = {q^n - p^n \over q-p^{-1}}\;\;\; ; \;\;\; q , p \epsilon C $ }

\vspace{1cm}
This case was discussed in \cite{mine2}. We have the following relations:
\begin{equation}
aa^+-qa^+a = p^{-N}
\end{equation}

The system of CS is now given by:
\begin{eqnarray}
||q,p,z> &=& \sum_{n\ge 0} {z^n\over \sqrt{[n]!}}|n> \nonumber \\
&=& \sum _{n\ge 0} {z^n \over [n]!}|0> \\
&=& exp_{q,p}^{(1)}(za^+)|0> \nonumber
\end{eqnarray}
where we have introduced the deformed exponential function $exp_{q,p}^{(1)}$.

\begin{itemize}
\item These states are eigenstates of the annihilation operator $a||q,p,z> = z||q,p,z>$

\item the states are also normalizable
\begin{eqnarray}
<q,p,z||q,p,z> &=& \sum_{n\ge 0}{|z|^{2n} \over \sqrt{\bar {[n]}![n]!}}\nonumber \\
&=& \sum _{n\ge 0} {|z|^2 \over |[n]|!} \\
&\equiv & exp_{q,p}^{(2)}(|z|^2) \nonumber
\end{eqnarray}
we introduced here another deformed exponential function\footnote{in the first case discussed in the previous section, we didn't need two deformed exponential functions since in this case we considered $q$ to be real, then $[n] = \bar {[n]}$} $exp_{q,p}^{(2)}(|z|^2)$.

The normalized states are thus given by:
\begin{eqnarray}
|q,p,z> &=& {\cal N}(|z|^2) ||q,p,z> \nonumber \\
&=& \bigg(exp_{q,p}^{(2)}(|z|^2)\bigg) ^{-{1\over 2}} exp_{q,p}^{(1)}(za^+)|0> 
\end{eqnarray}

\item The states are also continue in their label $z$, due to the continuity of the overlap term:
\begin{equation}
<q,p,z|q,p,z> = {\cal N}(|z|^2) {\cal N}(|z'|^2) exp_{q,p}^{(2)}(\bar z z')
\end{equation}

\item A resolution of unity in terms of the states is obtained in the spirit of (79)
\begin{equation}
\int \int _{|z|^2 < R_{q,p}} d^2z |q,p,z><q,p,z| W(|z|^2) = I
\end{equation}
where $R_{q,p}$ is the convergence radius of the two series defining $exp_{q,p}^{(1)}$ and $exp_{q,p}^{(2)}$; and is given by $ R_{q,p} = |q-p^{-1}| ^{-1}$

We have proved in \cite{mine2} that these series converges and that the resolution of unity is possible only in two cases\footnote{Otherwise either the series diverges or the resolution of unity is not possible}:

 - $|q| \le 1$ and $|p| =1$.

 - $|q|=1$ and $|p|\ge 1$

In these cases the weight function is given by:
\begin{equation}
W(x)= {{\cal N}^{(-2)}(x) \over 2\pi }\int_{-\infty}^{+\infty}e^{-iyx}\bar W(y)dy
\end{equation}
where$ \displaystyle \bar W(y) = \sum _{n\ge 0} {|[n]|!(iy)^n \over \pi n!}$

\end{itemize}

It is interesting to notice that, taught the case considered here is a particular case of the algebra ${\cal A}_q$, it generalizes at least three major known cases in the literature:
\begin{itemize}
\item when $p \longrightarrow q$ we obtain the Macfarlan case \cite{macfarlane}.

\item when $p \longrightarrow 1$ we obtain the conventional
quons \cite{greenberg}.

\item when $p \longrightarrow 1$ and $q \longrightarrow 1$ we obtain the
ordinary bosons.

\end{itemize}

With this we complete our discussion of the two approaches for the construction of CS for the deformed bosons. we consider now the case of fermions generalization.

\subsection{fermions deformations: k-fermions; the $Z_3$-graded case}

Before beginning with the generalization of the fermionic CS, we remind how things are done before deformations.

The fermionic HO is defined by:
\begin{equation}
\{a,a^+\} = aa^+ +a^+a =1 \;\;\;\; (a)^2 = (a^+)^2 =0
\end{equation}

Thus only two states are allowed $|0>$ and $|1>$. With these states we construct CS in the following way:
\begin{eqnarray}
|\xi > = e^{a^+\xi }|0> &=& (1 + a^+\xi )|0> \nonumber \\
&=& |0> - \xi |1>
\end{eqnarray}
where $\xi $ is a Grassmann variable. We recall that the Grassmann variables satisfy:
\begin{eqnarray}
\xi_1 \xi_2 &=&-\xi_2 \xi_1 \nonumber \\
\xi ^2 &=& 0 \\
\xi a &=& -a\xi \nonumber \\
\xi a^+ &=& -a^+ \xi \nonumber
\end{eqnarray}

The scalar product of two CS is given by:
\begin{equation}
<\bar {\xi}_1|\xi _2 > = 1+\bar {\xi}_1\xi _2
\end{equation}
where $\bar {\xi}$ is the dual of $\xi$, obeying the same relations (103)

The resolution of unity is obtained by:
\begin{equation}
\int d\bar {\xi}d\xi e^{-\bar {\xi} \xi}|\xi ><\bar{\xi} | = |0><0| + |1><1| = I
\end{equation}
where $e^{-\bar{\xi}\xi}= 1-\bar{\xi}\xi$

and the integrals are in the sens of Berezin \cite{berezin}:
\begin{equation}
\int d\xi = \int d\bar{\xi} =0 \;\;\;\;\; \int \xi d\xi = \int \bar{\xi}d\bar{\xi} = 1
\end{equation}

The generalization of the fermionic case is obtained as a particular case of the algebra ${\cal A}_q$ \cite{mine3}. In fact in ${\cal A}_q$, taking $\displaystyle [n] = {q^n-q^{-n} \over q-q^{-1}}$
we obtain the HO $aa^+ -qa^+a =q^{-N}$. Moreover when the parameter $q$ is a $k^{th}$ root of unity $ q= e^{2\pi i \over k}$, we can prove that $(a)^k = (a^+)^k =0$ which describes $k$-fermions. In what follows we consider the case $k=3$, and carry out the construction following\cite{mine3}.

The CS in this case are indeed given by:
\begin{eqnarray}
|\xi > &=& f(a^+\xi ) |0> \nonumber \\
&=& (1+ a^+\xi -  a^+\xi a^+\xi )|0> \\
&=& |0> + q^2\xi |1> - \sqrt{[2]} \xi^2 |2> \nonumber
\end{eqnarray}
where the $\xi$'s are $Z_3$-graded Grassman variables \cite{kerner1, kerner2}defined by:
\begin{equation} 
\xi_a\xi_b\xi_c = q\xi_b\xi_c\xi_a = q^2 \xi_c\xi_a\xi_b
\end{equation}
which implies:
\begin{eqnarray}
\xi^3 &=& 0 \\
\xi_a \xi_b\xi_c\xi_d &=& 0 \nonumber
\end{eqnarray}
with $a^+$ it behaves as follows:
\begin{equation}
\xi a^+ = q a^+ \xi
\end{equation}

No relations exists between $\xi a$ and $a\xi$, as well as for the binary products$\xi_a\xi_b$ and $\xi_b\xi _a$ which are considered to be independant.

We introduce $\bar {\xi}$ the dual of $\xi$. It obeys:
\begin{eqnarray}
\bar{\xi}_a\bar{\xi}_b\bar{\xi}_c &=& q^2 \bar{\xi}_b \bar{\xi}_c\bar{\xi}_a  \\
\bar {\xi}a &=& q^2 a \bar {\xi} \nonumber
\end{eqnarray}

The scalar product of two CS is given by:
\begin{eqnarray}
<\bar {\xi_1}|\xi_2> &=& 1 + q^2 \bar {\xi_1}\xi_2 - q\bar{\xi_1}\xi_2\bar{\xi_1}\xi_2 \\
&=& g(\bar{\xi_1}\xi_2) \nonumber
\end{eqnarray}
where
\begin{equation}
<\bar {\xi_1}| = <0| + q<1| \bar {\xi_1} - \sqrt{[2]} <2|\bar {\xi_1}^2
\end{equation}

A resolution of unity is given by:
\begin{equation}
\int d\bar{\xi} \; d\xi \; h(\bar{\xi}\xi) \; |\xi><\bar{\xi}| = |0><0| + |1><1| + |2><2|=I
\end{equation}
where:
\begin{equation}
h(\bar{\xi}\xi) = -q + \bar{\xi}\xi + \bar{\xi}\xi\bar{\xi}\xi
\end{equation}

and the integrals are computed using the Majid integrals \cite{majid} generalizing the Berezin ones.
\begin{eqnarray}
\int d\xi 1 &=& \int d\xi \;\xi = 0 \nonumber \\
\int d \bar{\xi} 1 &=& \int d \bar{\xi} \; \bar{\xi } = 0 \\
\int d\xi \xi^2 &=& \int d\bar {\xi}^2 \bar {\xi}^2 = 1 \nonumber
\end{eqnarray}

We have thus constructed CS for k-fermions (k=3). It is clear that using the same method CS related to k-fermions (with generic k) can be constructed for this we have to introduce the $Z_k$-graded Grassman variables.

We should mention that the q-deformed oscillators (bosonic or fermionic) related CS present many other analogies with the conventional CS. One of the most important properties is the fact that this deformed CS allow also the construction of an analogue of the Bargamann-Fock space: deformed Bargamann-Fock space for the first two cases considered in this section \cite{mine1} and a deformed Grassmann representative space for the k-fermions \cite{mine3}
\subsection*{$Z_3$-graded supercoherent states}

In addition to the parafermionic harmonic oscillator related to the $Z_3$-grassman algebra, we need a bosonic harmonic oscillator, in order to construct a $Z_3$-graded supersymmetric one.

The bosonic harmonic oscillator algebra is generated by the triplet $ \{M,b,b^+\}$, satisfying the usual commutation relations. 

\[
[b,b^+] = 1 \;\;\;, \;\;\; [b,M] = - b \;\;\; , \;\;\; [b^+,M] = b^+
\]
acting on the usual Fock space basis $\{ |m>, m=0,1,...\}$ as:

\[ 
b|m> = m |m> \;\;\;, \;\;\; b^+|m> = (m+1)|m+1> \;\;\;, \;\;\;M|m> = m|m>
\]

The bosonic coherent states is given for a complex $z$ by:

\begin{equation}
|z> = \sum _{m=0}^{\infty} {z^m \over {\sqrt{m!}}} |m>
\end{equation}

As in the case of $Z_2$-graded supersymmetry, a $Z_3$-graded supersymmetric coherent state is obtained by coupling this states with the states (13):

\begin{eqnarray}
|z,\xi > &=& |z> \otimes \; |\xi > \\
&=& D(z,\xi )\; |0>\otimes \; |0 > \nonumber
\end{eqnarray}
where $D(z,\xi) = e^{zb^+}f(\xi a^+)$

\section{conclusion}

The theory of deformation presents many interesting aspect and deserve thus to be studied at the deepest level. This was the case, and we can see now as a direct consequence of this, application of this theory in many areas of physics, other than the special physical systems for which it was first introduced. Perhaps the most interesting aspect of these applications, is the case of constructing particle models using non-commutative geometry-based models. But as mentioned in the introduction, no direct confrontation of these theories with experience is conceivable in the state of our current knowledge. This pushes us to introduce new tools in these theories. Due to their many intersting features coherent states are surely among these tools which should bring many solutions in this area. However, till now this axis is not sufficiently investigated.

We have thus given in this paper, a review on how things should be done for a particular case, namely CS related to q-deformed particles. We also presented a natural way of introducing these particles through teh deformation of the excterior algebra.

Still many aspects of these theories to be developed. One of these, is the relation of this deformed particles, to non-commutative theories. We believe that this theory (non-commutative), constitute the natural language in which a proper interpretation of these particles should be expressed.

\end{document}